\documentclass[epj]{svjour}
\usepackage{graphics}
\usepackage{latexsym}
\usepackage[latin1]{inputenc}
\usepackage{amsmath}
\usepackage{textcomp}
\usepackage{amsfonts}
\usepackage{amssymb}
\usepackage{mathrsfs}
\usepackage{color} 
\usepackage{graphicx}
\usepackage{subfigure}
\usepackage{epstopdf}
\usepackage{bm}
\usepackage{cite}

\begin{document}
\title{Cooperative behavior between oscillatory and excitable units:
  the peculiar role of positive coupling-frequency correlations}

\author{Bernard Sonnenschein\inst{1,2} \and Thomas K. DM.
  Peron\inst{3} \and Francisco A. Rodrigues\inst{4} \and J\"urgen
  Kurths\inst{1,2} \and Lutz Schimansky-Geier\inst{1,2}}

\institute{\inst{1} Department of Physics, Humboldt-Universit\"at zu
  Berlin, Newtonstrasse 15, 12489 Berlin, Germany\\\inst{2} Bernstein
  Center for Computational Neuroscience Berlin, Philippstrasse 13,
  10115 Berlin, Germany\\\inst{3} Instituto de F\'isica de S\~ao
  Carlos, Universidade de S\~ao Paulo, Avenida Trabalhador S\~ao
  Carlense 400, Caixa Postal 369, CEP 13560-970 S\~ao Carlos, S\~ao
  Paulo, Brazil\\\inst{4} Departamento de Matem\'atica Aplicada e
  Estat\'istica, Instituto de Ci\^encias Matem\'aticas e de
  Computa\c{c}\~ao, Universidade de S\~ao Paulo, Caixa Postal 668,
  13560-970 S\~ao Carlos, S\~ao Paulo, Brazil } 

\abstract{ We study the collective dynamics of noise-driven excitable
  elements, so-called active rotators.  Crucially here, the natural
  frequencies and the individual coupling strengths are drawn from
  some joint probability distribution. Combining a mean-field
  treatment with a Gaussian approximation allows us to find examples
  where the infinite-dimensional system is reduced to a few ordinary
  differential equations. Our focus lies in the cooperative behavior
  in a population consisting of two parts, where one is composed of
  excitable elements, while the other one contains only
  self-oscillatory units. Surprisingly, excitable behavior in the
  whole system sets in only if the excitable elements have a smaller
  coupling strength than the self-oscillating units.  In this way
  positive local correlations between natural frequencies and
  couplings shape the global behavior of mixed populations of
  excitable and oscillatory elements.
} 
\titlerunning{Cooperative behavior between oscillatory and excitable units}
\authorrunning{B. Sonnenschein \emph{et al.}}
\maketitle
\section{Introduction}
\label{intro}
Collective dynamics in biological systems is in general a complex
behavior that results from the interplay of non-identical, highly
nonlinear and noisy elements \cite{Gl01}.  Neuronal and cardiac
rhythms for instance originate from interactions among pacemaker and
excitable cells (see e.g. Refs.
\cite{Buz06,PoPoLSG12,LuBaSo13,LuBenBaBoTo14} and
\cite{DiFra93,ShaBoShGl13,Ka13,QuHuGarWe14}, respectively). Motivated
by these facts, we investigate the collective dynamics of coupled
non-identical elements, each being either excitable or
self-oscillatory. The latter shall model the pacemaking cells in
neuronal or cardiac tissues, for instance. Furthermore, in order to
make the model more realistic, individual coupling strengths are
allowed to be different. Our setting enables us to study how certain
correlations between the dynamics and the couplings on the microscopic
level affect the macroscopic behavior of the system. Many works
addressed the latter kind of question recently, see e.g.
Refs.~\cite{Br08,GogaGoArMo11,peron2012determination,leyva2012explosive,coutinho2013kuramoto,SonnSagSchi13,SkSuTaRes13,li2013reexamination,SuRuGuLi13,JiPeMeRodKu13,zhu2013criterion,ZhHuKuLiu13,LeNaSenAlmBuZaPaBoc13,zhu2013explosive,ChHHuShH13,ZoPerSmLiKu14,SkAr14}.
For interesting recent works that highlight the special interplay
between dynamics and network structure in neuronal systems, we refer
to \cite{mikkelsen2013emergence,orlandi2013noise}.

Of particular interest here are the works presented
in~\cite{ZhHuKuLiu13,ChHHuShH13}. Zhang \emph{et al.} considered
Kuramoto oscillators coupled in a generalized complex network.
Noteworthy, it was found that the crucial feature behind the emergence
of explosive synchronization\footnote{Explosive synchronization was
  coined by the finding of a discontinuous synchronization transition
  in scale-free networks of Kuramoto oscillators with bistability
  between incoherence and partial synchronization
  ~\cite{GogaGoArMo11}.} is a positive correlation between the natural
frequencies and the effective coupling strengths to the mean field
\cite{ZhHuKuLiu13}.  Chen \emph{et al.} studied effects of
degree-frequency correlations in a population of FitzHugh-Nagumo
neurons \cite{ChHHuShH13}. They extended in this way the finding of
explosive synchronization to relaxation oscillators with two separated
time scales.

The dynamical system that we study here puts emphasis on the
phenomenon of excitability, both on the local and the global scale.
Moreover, the coupling-frequency correlation considered in
\cite{ZhHuKuLiu13} shall motivate the specific formulation of our
model. To this end, we investigate the noise-driven active rotator
model introduced by Shinomoto and Kuramoto \cite{ShiKur86} with
distributed natural frequencies and coupling strengths. Specifically,
we analyze a system formed by two distinct parts of excitable and
self-oscillating units, the first having subthreshold natural
frequencies, while the other elements have frequencies above the
excitation threshold.

Based on previous findings and numerical observations it is reasonable
to approximate the phase distribution by a Gaussian with
time-dependent mean and variance
\cite{KurSchu95,ZaNeFeSch03,SonnZaNeiLSG13,SonnSchi13}.  Such an
assumption has also been made, e.g., for coupled FitzHugh-Nagumo
oscillators \cite{TaPak01,ZaSaLSGNei05}, integrate-and-fire neurons
\cite{Bu01}, a general class of master equations \cite{LafTo10} and
delayed-coupled systems
\cite{BuRanTodVas10,FraTodVasBu13,FraTodVasBu14}. Within the Gaussian
approximation the system's dimension can be reduced to four coupled
first-order differential equations, which allows a thorough
bifurcation analysis. On this basis, we distinguish the following
global states: (i) a resting state, where the units are silent, (ii) a
partially synchronized state, where a macroscopic fraction of the
units fire in synchrony and (iii) an incoherent state, where all the
units fire asynchronously. Finally, bistability between the resting
and the partially synchronized or the incoherent state is reported.

We find that heterogeneity both in the natural frequencies and the
coupling strengths impedes synchronization. However, a
counterintuitive phenomenon is found on top of this. A positive
coupling-frequency correlation where the self-oscillatory units
possess a stronger coupling than the excitable elements, brings the
whole system into an excitable state. Since individual coupling
strengths constrain how strongly single elements can feel the mean
field, the emergence of the excitable behavior on the global scale
crucially depends on how strongly the self-oscillating units are
influenced by the excitable elements.  Bistable behavior is only
found if the coupling-frequency correlation is sufficiently strong.

As an aside, our theory also yields analytical findings for the
stochastic Kuramoto model where temporal fluctuations act on the
frequencies and where the only source of quenched disorder is provided
by different coupling strengths. We discover that the mean-field
amplitude of the oscillators with weaker coupling can scale
anomalously in dependence on the average coupling strength, giving
rise to chimera-like states (see appendix \ref{KuramotoCouplings}).

The paper is organized as follows: In Sec. \ref{model} we present the
model and explain its basic properties. Section \ref{sec_MF} is
devoted to the mean-field treatment and to the derivation of the
Fokker-Planck equation for the probability distribution of the phases.
In Sec. \ref{gaussian_approx} we prepare the low-dimensional behavior
of the model through the Gaussian approximation technique, and in Sec.
\ref{example} we work out a specific example. Numerical results
thereby corroborate the theoretical findings. Final conclusions are
drawn in Sec. \ref{conclusion}.

\section{Model}
\label{model}
Consider a population of noise-driven active rotators \cite{ShiKur86},
where the dynamics of individual phases $\phi_i(t)$ follows
\begin{equation}
  \dot{\phi}_i=\omega_i-a\sin\phi_i+\frac{K_i}{N}\sum_{j=1}^{N}\sin\left(\phi_j-\phi_i\right)+\xi_i(t).
\label{ourmodel}
\end{equation}
The units are indexed by $i=1,\ldots,N$.
The parameter $a$, which determines the excitation threshold, is the same for all rotators.
Natural frequencies are
denoted by $\omega_i$. Each element is coupled to
the others with an individual strength, $K_i$. We
  will assume that the individual frequencies and coupling strengths
are random numbers that are drawn from the same joint probability distribution $P(\omega,K)$,
independently between the elements.
   In addition, we assume that the initial phases
  of the active rotators $\phi_i^0$ at the starting time $t^0$ are
  independent and given by a distribution density $P_{\mathrm{in}}(\phi^0)$.

We emphasize that the values for the $\omega$'s and $K$'s
 are chosen initially and then stay fixed during the
  whole evolution of the system. They represent frozen random
  variables (``quenched disorder''), which shall be some real numbers.
  We do not consider repulsive interactions here, that is the coupling
  strengths are non-negative.

The terms $\xi_i(t)$ shall model the accumulated effect of various
sources of temporal fluctuations that may result from a noisy
environment, cell-intrinsic noise and stochasticity in the
interactions. Lumped together, one may assume zero mean Gaussian white
noise sources \cite{AniAstNeVaSchi07,LiGarNeiSchi04}. Then one has
\begin{equation}
  \label{eq:nooise}
  \langle\xi_i(t)\rangle=0, ~~~~\langle\xi_i(t)\xi_j(t')\rangle=2D\delta_{ij}\delta\left(t-t'\right),  
\end{equation}
where the second relation expresses the lack of memory in the noise
and that noise at one element is independent from the other ones. The
angular brackets denote averages over different realizations of the
noise and $D$ is the noise intensity.

For an isolated element without additive noise,
$\dot{\phi}_i=\omega_i-a\sin\left(\phi_i\right)$, the excitable
behavior is easily understood. For $\left|a\right|>\omega_i$ the
stable equilibrium is located at
$\phi_i^{\infty}=\arcsin(\omega_i/a)$, and the unit needs a
sufficiently strong perturbation in order to make a big excursion.
Noise can play this role driving the system to escape from the resting
state $\phi_i^{\infty}$. An escape event corresponds to the release of
a single spike \cite{ShiKur86}.  For $\left|a\right|<\omega_i$ the
element shows oscillatory behavior with frequency
$\sqrt{\omega_i^2-a^2}$. It has to be emphasized that the phase
variable $\phi$ does not rotate uniformly; it is slowest near
$\phi=\pi/2$ and fastest near $\phi=3\pi/2$.

\section{Mean-field theory}
\label{sec_MF}
Instead of the Langevin equations \eqref{ourmodel} the system of $N$
coupled active rotators can be described by the joint probability density
\begin{equation}
  \label{eq:npdf}
\mathcal{P}_N\left(\boldsymbol{\phi},t;\boldsymbol{\phi}^0,t^0;\boldsymbol{\omega},\boldsymbol{K}\right).
\end{equation}
The vector $\boldsymbol{\phi}=\left(\phi_1,\ldots,\phi_N\right)$ is built from the phases of $N$ rotators at time
$t$, and $\boldsymbol{\phi}^0=\left(\phi_1^0,\ldots,\phi_N^0\right)$ consists
of their values at the initial time $t^0$.  The time-independent vectors
$\boldsymbol{\omega}=\left(\omega_1,\ldots,\omega_N\right)$ and $\boldsymbol{K}=\left(K_1,\ldots,
  K_N\right)$ are composed of, respectively, the natural frequencies
and the coupling strengths of the $N$ rotators. Normalization requires
\begin{equation}
\begin{aligned}
  \label{eq:norm}
\int_{0}^{2\pi} \mathrm d^N{\boldsymbol{\phi}} \int_{0}^{2\pi} \mathrm d^N{\boldsymbol{\phi}^0} \int_{-\infty}^{+\infty} \mathrm d^N{\boldsymbol{\omega}} \int_{0}^{+\infty} \mathrm d^N{\boldsymbol{K}}\,\mathcal{P}_N \,= \,1\,.
\end{aligned}
\end{equation}
For rotators with given $\boldsymbol{\omega}$ and $\boldsymbol{K}$, this joint probability distribution is related with the conditional
probability density
${p}_N\left(\boldsymbol{\phi},t|\boldsymbol{\phi}^0,t^0;\boldsymbol{\omega},\boldsymbol{K}\right)$
from the initial state $\boldsymbol{\phi}^0$ at $t^0$ to the present
phases $\boldsymbol{\phi}$ at time $t$ as
\begin{equation}
\begin{aligned}
  \label{eq:trans}
&\mathcal{P}_N\left(\boldsymbol{\phi},t;\boldsymbol{\phi}^0,t^0;\boldsymbol{\omega},\boldsymbol{K}\right) \,=\\
&\ \ \ \ \ \ {p}_N\left(\boldsymbol{\phi},t|\boldsymbol{\phi}^0,t^0;\boldsymbol{\omega},\boldsymbol{K}\right)\,{P}_N\left(\boldsymbol{\phi}^0,t^0;\boldsymbol{\omega},\boldsymbol{K}\right)\,.
\end{aligned}
\end{equation}
By assuming independent initial phases and pairs $\omega_i$ and
$K_i$ at the nodes, we can factorize
\begin{equation}
  \label{eq:joint2}
  {P}_N\left(\boldsymbol{\phi}^0,t^0;\boldsymbol{\omega},\boldsymbol{K}\right)= \prod_{i=1}^N \,P_{\mathrm{in}}(\phi_i^0)\,P(\omega_i, K_i)\,. 
\end{equation}

The joint probability density is governed by a linear Fokker-Planck
equation (FPE) which describes the evolution of the population from
time $t^0$ to time $t>t^0$ \cite{Ris96}:
\begin{equation}
\begin{aligned}
  \frac{\partial\mathcal P_N}{\partial t}=&\ D\sum_{i=1}^N\frac{\partial^2\mathcal P_N}{\partial\phi_i^2}-\sum_{i=1}^N\frac{\partial}{\partial\phi_i}\mathcal P_N\times\\
  &\times\left[\omega_i-a\sin\phi_i+\frac{K_i}{N}\sum_{j=1}^{N}\sin\left(\phi_j-\phi_i\right)\right].
  \label{FPE_N}
\end{aligned}
\end{equation}
The usual way to proceed is to introduce reduced probability densities
$\mathcal P_n$ with index $n=1,2,\ldots,N-1$ by integrating $\mathcal
P_N$ over a subset of variables and parameters. Since all rotators
are identical in their dynamic behavior with respect to the specific
frequencies and coupling constants, we take exemplarily rotators
with labels $i=1,\ldots,n$ and integrate over respective
variables and parameters with numbers $i>n$. This defines the reduced
probability densities with integration boundaries as in \eqref{eq:norm}:
\begin{equation}
\begin{aligned}
\label{one_part}
  \mathcal P_n &\left(\phi_1,t; \phi_1^0,t^0;\omega_1,K_1;\ldots;\phi_n,t; \phi_n^0,t^0;\omega_n,K_n\right)\,=\\
&\nonumber \int \prod_{i=n+1}^N \left(\mathrm d{\phi_i} \mathrm d{\phi_i^0} \mathrm d{\omega_i} \mathrm d{K_i}\right)\,\,\mathcal{P}_N \left(\boldsymbol{\phi},t;\boldsymbol{\phi}^0,t^0;\boldsymbol{\omega},\boldsymbol{K}\right)\,.
\end{aligned}
\end{equation}
To obtain the dynamics for these densities, we integrate the FPE \eqref{FPE_N} over
the corresponding subset of the variables and the other quantities.
Then one is left with a set of coupled differential equations, akin to
a Bogoliubov-Born-Green-Kirkwood-Yvon (BBGKY) hierarchy.  Truncating
this hierarchy at some $n$ leads to a reduced description.

Specifically here, we will be interested in the one-osc\-il\-la\-tor probability 
density $\mathcal P_1$. Therefore, we integrate the FPE \eqref{FPE_N} over the $N-1$ phases
$\phi_2,\ldots,\phi_N$, their initial values $\phi_2^0,\ldots,\phi_N^0$, the natural
frequencies $\omega_2,\ldots,\omega_N$ and the coupling strengths
$K_2,\ldots,K_N$. This yields
\begin{equation}
\begin{aligned}
  &\frac{\partial\mathcal P_1}{\partial t}=D\frac{\partial^2\mathcal P_1}{\partial\phi_1^2}-\frac{\partial}{\partial \phi_1}\left(\omega_1-a\sin\phi_1\right)\mathcal P_1 \\
  &\ -\frac{K_1(N-1)}{N}\frac{\partial}{\partial\phi_1}\int \mathrm d \phi_2 \int \mathrm d \phi^0_2 \int \mathrm d \omega_2 \int \mathrm d K_2 \,\times\\
&\times \, \sin\left(\phi_2-\phi_1\right)\mathcal
  P_2\left(\phi_1,\phi_2,t;\phi^0_1,\phi^0_2,t^0;\omega_1,K_1,\omega_2,K_2\right).\\
\end{aligned}
  \label{two-osc}
\end{equation}
Note that $\mathcal P_1(\phi_1,t;\phi_1^0,t^0;\omega_1,K_1)$ relates hierarchically to $\mathcal P_2(\phi_1,\phi_2,t;\phi_1^0,\phi_2^0,t^0;\omega_1,K_1,\omega_2,K_2)$ being 
the two-osc\-ill\-ator distribution.

The investigation of coupling-coupling or frequency-frequency
correlations shall remain a topic for future research. However, we
will allow dependencies between the natural frequency and the
coupling strength at each node, given by the joint
distribution $P(\omega,K)$. Having this in mind, we will assume that the dynamical
correlations between the phases of two arbitrarily chosen oscillators
can be discarded as follows:
\begin{equation}
\begin{aligned}
&\mathcal P_2\left(\phi_1,\phi_2,t;\phi_1^0,\phi_2^0,t^0;\omega_1,K_1,\omega_2,K_2\right)\equiv \\
&\ \mathcal P_1\left(\phi_1,t;\phi_1^0,t^0;\omega_1,K_1\right)\mathcal P_1\left(\phi_2,t;\phi_2^0,t^0;\omega_2,K_2\right).
\label{propchaos}
\end{aligned}
\end{equation} 
This corresponds essentially to the lowest-order truncation of the BBGKY
hierarchy. In particular, Eq. \eqref{two-osc} becomes closed but
nonlinear in $\mathcal P_1$.  Remarkably, in
the thermodynamic limit of infinitely many elements,
$N\rightarrow\infty$, such a truncation can be justified in a rigorous
way for various systems (for recent overviews along with new results,
cf. Refs.  \cite{faugeras2009constructive,mischler2013kac}). The
argument goes back to Boltzmann's ``Stosszahlansatz'', which was later
rigorously formalized by Kac with the concept of ``propagation of
molecular chaos'' \cite{kac1954foundations}. In the light of those
achievements (see also \cite{bonilla1993glassy,HilBuCh07} and
references therein), \eqref{propchaos} can be considered to be exact in the
thermodynamic limit.

Henceforth we neglect the indices at $\phi,\ \omega$ and $K$,
as the underlying assumption in the mean-field approach is that rotators with 
the same natural frequency and coupling strength are statistically identical.
Moreover, we proceed with the conditional form of the one-oscillator
probability density $p_1$, which is obtained from $p_N$ [see Eq. \eqref{eq:trans}] 
after appropriate integration \cite{SonnSchi12}. 
For every given pair $(\omega,K)$, the
expression $p_1\left(\phi,t|\phi^0,t^0;\omega,K\right)\mathrm d\phi$
denotes the fraction of oscillators, which start with the phase $\phi^0$ at time $t^0$ and then have
a phase value between $\phi$ and $\phi+\mathrm d\phi$ at time $t$. It follows from integration of the
$p_N$ and from an average over frequency and
coupling constants of the other units of the ensemble.  Accordingly,
the normalization $1=\int_0^{2\pi}\mathrm d\phi\ p_1\left(\phi,t|\phi^0,t^0;\omega,K\right)\ \forall\ \phi^0,\omega,K$ must be satisfied.

For the dynamical evolution of
$p_1\left(\phi,t|\phi^0,t^0;\omega,K\right)$ one gets the following 
nonlinear Fokker-Planck \cite{Fr05} or McKean-Vlasov equation:
\begin{equation}
  \label{fpe}
  \begin{aligned}
    \frac{\partial p_1 }{\partial
      t}\,=&\,-\,\frac{\partial}{\partial\phi} v_{\omega,K}(\phi,t)\,
    p_1 +D\frac{\partial^2 p_1}{\partial\phi^2}.
 \end{aligned}
\end{equation}
Nonlinearity enters equation~\eqref{fpe} through the mean increment of
the phase per unit time, i.e.
\begin{equation}
  v_{\omega,K}(\phi,t)\equiv\omega-a\sin\phi+r K\sin\left(\Theta-\phi\right),
\end{equation}
which depends on the density $p_1$ via the mean-field amplitude $r(t)$
and phase $\Theta(t)$,
\begin{equation}
  r(t)\mathrm e^{i\Theta(t)}=\langle\langle r_{\omega',K'}(t)\ \mathrm e^{i\Theta_{\omega',K'}(t)}\rangle\rangle.
  \label{order_fpe}
\end{equation}
The averages $\langle\langle\ldots\rangle\rangle\equiv\int\mathrm
d\omega'\int\mathrm dK'\ldots P(\omega',K')$ connect in a superposed
manner the global with the following local mean-field variables,
\begin{equation}
\begin{aligned}
&  r_{\omega,K}(t)\mathrm e^{i\Theta_{\omega,K}(t)}=\\
&\,\,\,\,\int_{0}^{2\pi}\mathrm d\phi\int_{0}^{2\pi}\mathrm d\phi^0\mathrm e^{i\phi} p_1\left(\phi,t|\phi^0,t^0;\omega,K\right) P_{\mathrm{in}}(\phi^0).
  \label{localorder}
\end{aligned}
\end{equation}
The set of equations \eqref{fpe}-\eqref{localorder} has to be solved
with the initial condition for the transition probability density
\begin{equation}
  \label{eq:init_trans}
  p_1\left(\phi,t^0|\phi^0,t^0;\omega,K\right)=\delta\left( \phi-\phi^0\right)\,,\,\forall\ \omega,K\,. 
\end{equation}

Alternatively, we can formulate the problem in terms of a nonlinear FPE for
the marginal density of the phase $\phi$ at time $t$,
\begin{equation}
  \label{eq:marg}
  p_1\left(\phi,t|\omega,K\right)\,=\,\int_0^ {2\pi} \mathrm d \phi^ 0 p_1\left(\phi,t|\phi^0,t^0;\omega,K\right) P_{\mathrm{in}}(\phi^ 0)\,.
\end{equation}
Specifically, this marginal density replaces the
conditional probability density in the nonlinear FPE \eqref{fpe}
and in Eq. \eqref{localorder} via integration over the initial
phases. Then we have an initial value problem that has to be solved,
in agreement with former assumptions, 
with the initial condition 
\begin{equation}
  \label{eq:init}
  p_1\left(\phi,t^0|\omega,K\right)\,=\,P_{\mathrm{in}}(\phi^0)\,,\,\forall\ \omega,K\,.
\end{equation}
In the derivation of \eqref{fpe}-\eqref{localorder}, we did not drop the dependence of the conditional
probability density on the initial state. Since the FPE is nonlinear,
the temporal evolution of the mean field and of the drift term in
\eqref{fpe} can sensitively depend on the initial distribution of the
phases. We also note that the assumption of propagation of chaos
appears to be problematic in sparsely connected networks, cf. section
VII. in Ref.  \cite{SonnSchi12} for numerical findings.

We remark that the nonlinear FPE \eqref{fpe} comprises a
large system of coupled partial differential equations. The nodes with
coinciding pairs of frequency and coupling constants can be
interpreted as one species. Every species obeys the FPE with the
corresponding $\omega$ and $K$. They contribute with subfields given by Eq.
\eqref{localorder} in accordance with their emergence to the mean field
\eqref{order_fpe}. The latter is given by the probability density
$P(\omega,K)$ about which the subfields are averaged.

\section{Fourier series expansion and Gaussian approximation}
\label{gaussian_approx}
We proceed to study the evolution of the marginal density $p_1\left(\phi,t|\omega,K\right)$, Eq. \eqref{eq:marg}. 
Since it is $2\pi$-periodic in $\phi$, we can first write a Fourier series expansion:
\begin{equation}
p_1(\phi,t|\omega,K)=\frac{1}{2\pi}\sum_{n=-\infty}^{+\infty}\rho_n(t|\omega,K)\mathrm e^{-in\phi}\ ,
\label{expansion} 
\end{equation}
with $\rho_0=1$ and $\rho_{-n}=\rho_{n}^*$.

Inserting \eqref{expansion} into \eqref{fpe}, multiplying by $\exp(im\phi),$ $m\in\mathbb Z$ and collecting the non-zero terms after integration over $\phi$, one obtains an infinite chain of coupled complex-valued differential equations for the Fourier coefficients $\rho_n(t|\omega,K)$. That is, for every pair $\left(\omega,K\right)$ we can write
\begin{equation}
\begin{aligned}
\frac{\dot\rho_n(t|\omega,K)}{n}=&\frac{a}{2}\Bigl[\rho_{n-1}(t|\omega,K)-\rho_{n+1}(t|\omega,K)\Bigr]\\
&\ -(Dn-i\omega)\rho_{n}(t|\omega,K)\\
&\ +\frac{K}{2}\left[\rho_{n-1}(t|\omega,K)\langle\langle\rho_1(t|\omega',K')\rangle\rangle\right.\\
&\ \left.-\rho_{n+1}(t|\omega,K)\langle\langle\rho_{-1}(t|\omega',K')\rangle\rangle\right].
\end{aligned}
\label{chain}
\end{equation}
An additional average appears if one considers complex networks in a coarse-grained way \cite{SonnZaNeiLSG13}.
While \eqref{chain} provides an exact representation of the system, it is not possible to derive the solutions
in an explicit way due to its hierarchical character. Since the Fourier coefficients rapidly decay with growing $n$,
one can get accurate results by truncating the hierarchy at a large enough $n$. Here we aim for an approximate
dimensionality reduction that allows bifurcation analysis or even explicit solutions in important limiting cases. 
This is the topic of the next sections.

We first seek a closure of the infinite set of equations \eqref{chain}. 
The Ott-Antonsen ansatz \cite{OttAnt08} achieves this in an exact manner for deterministic ensembles of coupled phase oscillators. Unfortunately, for the case with temporal fluctuations the direct application of the Ott-Antonsen ansatz is not possible, and we are unaware of its appropriate modifications.

Here we use instead a Gaussian approximation (GA): we assume that in every subset of oscillators with the same individual quantities $\left(\omega,K\right)$, the distribution of the phases at every moment of time is Gaussian with mean $m_{\omega,K}(t)$ and variance $\sigma_{\omega,K}^2(t)$ \cite{ZaNeFeSch03,SonnZaNeiLSG13}. 

Consider separately the real and imaginary parts of the Fourier coefficients \eqref{expansion}, that is 
\begin{equation}
  \label{eq:ampl}
  \rho_n(t|\omega,K)\equiv c_n(t|\omega,K)+is_n(t|\omega,K)\,.  
\end{equation}
In the thermodynamic limit $N\to\infty$ the GA then yields
\begin{equation}
\begin{aligned}
c_n(t|\omega,K)&=\exp\left[-n^2\sigma_{\omega,K}^2(t)/2\right]\cos\left[nm_{\omega,K}(t)\right],\\
s_n(t|\omega,K)&=\exp\left[-n^2\sigma_{\omega,K}^2(t)/2\right]\sin\left[nm_{\omega,K}(t)\right].
\label{cs}
\end{aligned}
\end{equation}
As a result, all $c_n$ and $s_n$ are given as combinations of $c_1$ and $s_1$: $c_2=c_1^4-s_1^4$, $s_2=2s_1c_1\left(s_1^2+c_1^2\right)$, etc. \cite{ZaNeFeSch03}.

By transforming the variables $\left\{c_1(t|\omega,K), s_1(t|\omega,K)\right\}$ to the first two cumulants of the Gaussian distribution, $\left\{m_{\omega,K}(t), \sigma^2_{\omega,K}(t)\right\}$, we obtain the following pair of differential equations:
\begin{equation}
\begin{cases}
&\dot m_{\omega,K}=\omega-\exp\left(-\sigma_{\omega,K}^2/2\right)\cosh\sigma_{\omega,K}^2 \left[a\sin m_{\omega,K}\right.\\
&\ \ -\left.K\Big\langle\Big\langle\exp\left(-\sigma_{\omega',K'}^2/2\right)\sin\left(m_{\omega',K'}-m_{\omega,K}\right)\Big\rangle\Big\rangle\right],\\
&\dot{\sigma}^2_{\omega,K}/2=D-\exp\left(-\sigma_{\omega,K}^2/2\right)\sinh\sigma_{\omega,K}^2 \left[a\cos m_{\omega,K}\right.\\
&\ \ +\left.K\Big\langle\Big\langle\exp\left(-\sigma_{\omega',K'}^2/2\right)\cos\left(m_{\omega',K'}-m_{\omega,K}\right)\Big\rangle\Big\rangle\right].
\end{cases}
\label{dotmvar}
\end{equation}
Thus, for a continuous coupling-frequency distribution $P(\omega,K)$ the reduced system is 
still infinite-dimensional, because for any pair $\left(\omega,K\right)$ one has to solve the two 
differential equations \eqref{dotmvar}, and all of those are coupled through the averages $\langle\langle\ldots\rangle\rangle$. 
In order to obtain a low-dimensional system, we need to continue with a discrete coupling-frequency distribution $P(\omega,K)$ with a finite number of 
different $\omega$'s and $K$'s. Indeed, interesting example systems are readily found, as shown in the next section.
 
Before coming to the integral part of our analysis, we would like to mention that one can also perform a variable transformation to the local mean-field variables:
\begin{equation}
\begin{cases}
\dot{r}_{\omega,K}=&-r_{\omega,K}D+\frac{1-r_{\omega,K}^4}{2}\left[a\cos\Theta_{\omega,K}\right.\\
&+\left.K\Big\langle\Big\langle r_{\omega',K'}\cos\left(\Theta_{\omega',K'}-\Theta_{\omega,K}\right)\Big\rangle\Big\rangle\right],\\
\dot{\Theta}_{\omega,K}=&\omega-\frac{r_{\omega,K}^{-1}+r_{\omega,K}^3}{2}\left[a\sin\Theta_{\omega,K}\right.\\
&-\left.K\Big\langle\Big\langle r_{\omega',K'}\sin\left(\Theta_{\omega',K'}-\Theta_{\omega,K}\right)\Big\rangle\Big\rangle\right].
\end{cases}
\label{dotrtheta}
\end{equation}
Note that the mean phases are not defined in the case of vanishing mean-field amplitudes.

Let us briefly recapitulate. We have derived an approximate lower-dimensional description for the infinite-dimensional system \eqref{ourmodel}, expressed by $\dot{c_1}(t|\omega,K), \dot{s_1}(t|\omega,K)$ or
$\dot{m}_{\omega,K}(t), \dot{\sigma}^2_{\omega,K}(t)$ or $\dot{r}_{\omega,K}(t), \dot{\Theta}_{\omega,K}(t)$. 
All three descriptions are equivalent. Now one could directly plug in a correlation between the coupling strength $K$ and the natural frequency $\omega$ as proposed by Zhang \emph{et al.} \cite{ZhHuKuLiu13}. Many interesting examples are conceivable, and we expect further fruitful investigations building on what we have just derived.

In the past, understanding the effects of heterogeneity benefited immensely by dividing the whole system into two subpopulations, see e.g. Refs. \cite{SonnZaNeiLSG13,MoKuBl04,AbMiStWi08,La09,HoStr11PRL,HoStr11,PaMo14}.
This strategy will also be adopted in the next section.

\section{Mixed population of excitable and self-oscillatory elements}
\label{example}
On the basis of the reduced description derived in the previous section, we will now turn our attention to an interesting example
\begin{figure*}
\centering
\includegraphics[width=0.85\linewidth]{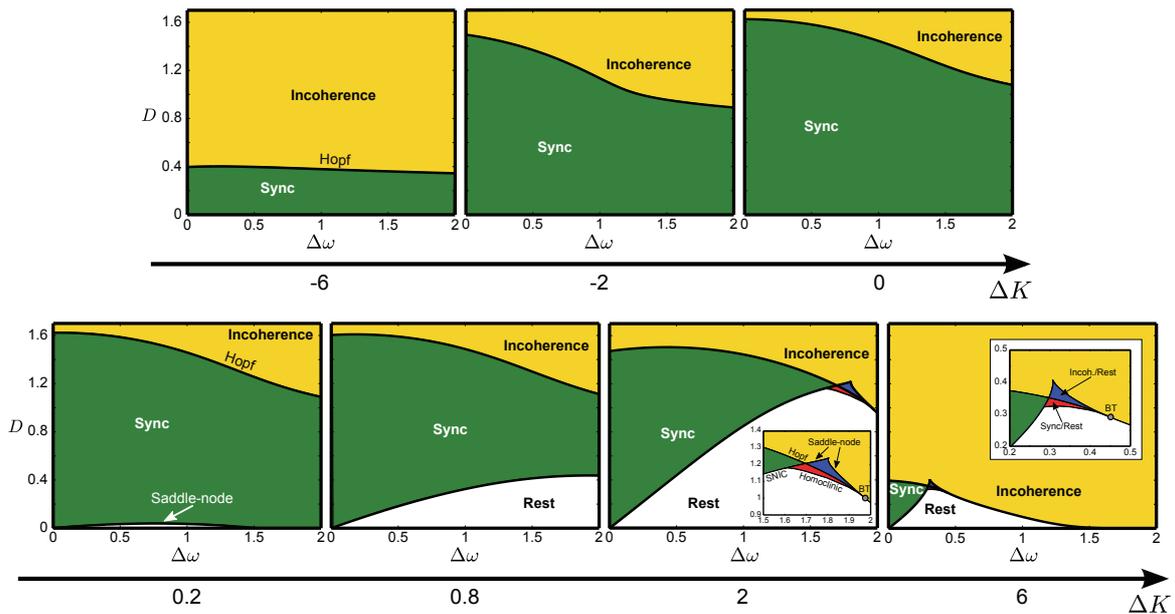}
\caption[]{(Color online) Hopf and saddle-node bifurcations in the plane spanned by the noise intensity $D$ and the frequency mismatch $\Delta\omega$, as obtained from a bifurcation analysis of the
reduced system \eqref{mvar}. Changes in the bifurcation diagram are shown as a function of the coupling mismatch $\Delta K$, with a fixed average coupling strength of $K_0=4$. Green shaded areas with 
label ``sync" represent the partially synchronized state, while in the yellow shaded regions no synchronized oscillations are found. In the white areas the system is at rest, in which the excitable 
units do not fire. For $\Delta K=2$ and $\Delta K=6$ insets show in more detail the parameter regions that correspond to bistable dynamics. The latter are found between the saddle-node bifurcation 
curves and the homoclinic bifurcation line which emanates from a Bogdanov-Takens bifurcation (BT). Moreover, the bistable region is separated into two parts by the Hopf bifurcation line. In the red
area below the Hopf line partially synchronized and resting state coexist, while in the blue area resting and incoherence coexist.}
\label{bif}
\end{figure*}
that allows a detailed analysis. We consider a mixed population consisting of two equally sized constituents; one half is chosen to be excitable and the other half shall be self-oscillating. 
This is realized by choosing one natural frequency below and the other one above the excitation threshold. Furthermore, both 
subpopulations shall have their own coupling strengths. 
Hence, for the coupling-frequency distribution we take a sum of two delta functions,
$P(\omega,K)=p\delta\left[\left(\omega,K\right)-\left(\omega_1,K_1\right)\right]+(1-p)\delta\left[\left(\omega,K\right)-\left(\omega_2,K_2\right)\right]$,
with $p=0.5$, $0<\omega_1<1$, $\omega_2>1$, and $K_{1,2}>0$.
In particular, we proceed with the following four-dimensional system [cf. Eq. \eqref{dotmvar}]:
\begin{equation}
 \begin{cases}
\dot m_1&=\omega_1-\mathrm e^{-\sigma_1^2/2}\cosh \sigma_1^2 \left[\sin m_1\right.\\
&+\left.(K_1/2)\ \mathrm e^{-\sigma_2^2/2}\sin(m_1-m_2)\right], \\
\dot \sigma_1^2/2&=D-\mathrm e^{-\sigma_1^2/2}\sinh \sigma_1^2 \left\{\cos m_1\right.\\
&+\left.(K_1/2)\left[\mathrm e^{-\sigma_1^2/2}+\mathrm e^{-\sigma_2^2/2}\cos(m_1-m_2)\right]\right\}, 
 \end{cases}
\label{mvar}
\end{equation}
where $\omega_1=1-\Delta\omega/2$ and $K_1=K_0-\Delta K/2$. The equations for $\dot m_2$ and $\dot \sigma_2^2$ are similar; just interchange $1$'s with $2$'s, 
and set $\omega_2=1+\Delta\omega/2$, $K_2=K_0+\Delta K/2$. 
Indices $i=1,2$ are abbreviations for $\left\{\omega_i,K_i\right\}$.
Henceforth, we call the differences in the natural frequencies and coupling strengths \emph{frequency mismatch} ($\Delta\omega$) and \emph{coupling mismatch} ($\Delta K$), respectively.
Note that the above choice is such that the average frequency and coupling strength are not affected by the mismatches.
Equations for the four mean-field variables follow \textit{mutatis mutandis} from Eq. \eqref{dotrtheta}.

Similar problems were addressed in the context of oscillatory systems, where parts are inactivated due to aging \cite{DaNak04,PaMon06,DaKaNi13}. Another mixed population of excitable and ``driver" units was studied by Alonso and Mindlin \cite{AlMi11}. Finally, the recent work \cite{LuBaSo13} puts forward a detailed analysis of coupled theta neurons where both inherently spiking and excitable neurons are present.

From now on we study the collective behavior in system \eqref{mvar} with the help of MATCONT \cite{DhGoKuz03}, namely in dependence on four dimensionless parameters: 
the noise intensity $D$, the frequency mismatch $\Delta\omega$, the coupling mismatch $\Delta K$, and the average coupling strength $K_0$.

The frequency mismatch $\Delta\omega$ is varied in the interval $\left(0,2\right)$,
restricting to positive natural frequencies. 
The coupling mismatch $\Delta K$ can take values between $\left(-2K_0,2K_0\right)$. Positive (negative) values of $\Delta K$ can be referred to as positive (negative)
coupling-frequency correlations, as long as there is a frequency mismatch $\Delta\omega>0$.
Similar as in \cite{TesScToCo07}, we focus here first on $K_0=4$; in the appendix \ref{accuracy} we show results for smaller average coupling strengths.

Coupled excitable elements stay at rest, if they cannot globally surpass the excitation threshold. If they do, the question then is whether a macroscopic fraction of them fires in synchrony, which amounts to 
a partially synchronized state, or whether the firing is completely incoherent among the elements.
Fig. \ref{bif} depicts the Hopf and saddle-node bifurcations that delineate those three states. Additional Hopf and saddle-node bifurcations that come after unstable equilibria can be neglected, because they do not affect the dynamics.
\begin{figure}
\centering
\includegraphics[width=0.95\linewidth]{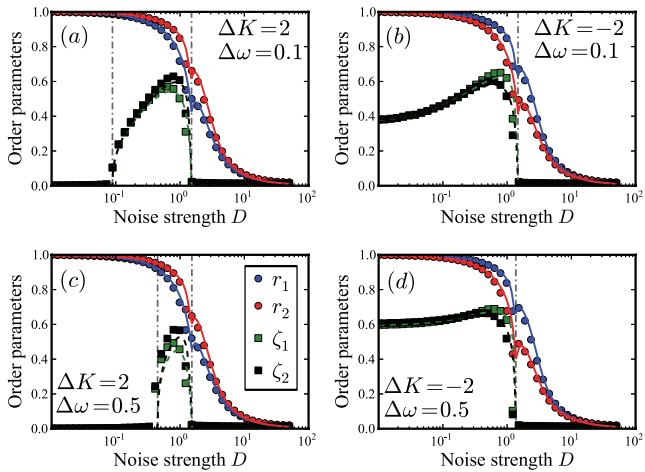}
\caption[]{(Color online) Long-time averaged Kuramoto ($r_{1,2}$) and Kuramoto-Shinomoto order parameters ($\zeta_{1,2}$, cf. Eq. \eqref{zeta1}), theory (lines) vs. simulation (dots). 
Vertical dash-dotted lines correspond to Hopf and SNIC bifurcations, respectively, as it can be
extracted from Fig. \ref{bif}. The average coupling strength is fixed at $K_0=4$.}
\label{test0}
\end{figure}
As one would expect, increased frequency and coupling mismatches impede the emergence of collectively synchronized oscillations. Specifically, above the Hopf bifurcation line the oscillatory units fire
incoherently, while below the Hopf line a synchronized firing sets in. Interestingly, a positive coupling-frequency correlation gives rise to 
a qualitative change in the global dynamics, since the saddle-node bifurcation shows up for $\Delta K>0$. Indeed the critical value for this phenomenon is found to equal $\Delta K_c=0$. This is 
visualized in Fig. \ref{bif} for a small coupling mismatch of $\Delta K=0.2$. We note that for $\Delta\omega\rightarrow 0$ the saddle-node bifurcation line always goes to vanishing noise intensity $D=0$, 
independently of $\Delta K$. Below the saddle-node curve, the excitable elements are resting and do not fire. For increasing $\Delta K$, the saddle-node line bends upwards, culminating in a Bogdanov-Takens bifurcation (BT),
which is located at an intersection of the Hopf and the saddle-node lines. From the BT a homoclinic bifurcation line emanates, which ultimately merges with the saddle-node curve 
(then called a SNIC bifurcation line), see the insets in Fig. \ref{bif}. We calculate the homoclinic bifurcations as follows. Starting at the BT we continue the Hopf bifurcation for some time steps, then switch 
the continuation to the limit cycle while tracking the period with the noise intensity $D$ as the control parameter. At the homoclinic bifurcation the period of the limit cycle diverges. We accept the $D$ values if they do not change anymore in the order of $10^{-4}$ upon approaching the divergence. The whole procedure is repeated until the homoclinic bifurcation line reaches the saddle-node curve. Both for $\Delta K=2$ and $\Delta K=6$ we save hereby in total eight pairs of $\left(\Delta\omega, D\right)$ and connect them by a line, see Fig. \ref{bif}.

Importantly, the area between the saddle-node bifurcations and the homoclinic bifurcation line corresponds to bistable (hysteretic) dynamics. In particular, two qualitatively different bistable
 dynamics are separated by the Hopf bifurcation line; below it, the resting and the partially synchronized state coexist, whereas above there is a coexistence between two steady states, the resting
and the incoherent state (compare with Ref. \cite{ZaNeFeSch03}).

Besides performing a bifurcation analysis, another way of characterizing the collective dynamics lies in calculating suitable order parameters. One of them is the classical Kuramoto
order parameter, Eqs. \eqref{order_fpe}, \eqref{localorder}, which measures how similar the phase variables are to each other. However, it is not sufficient here to consider this order parameter, because in case 
of slowly varying phases, it would attain large values \cite{ShiKur86}. In the extreme case of resting elements, the Kuramoto order parameter would be even equal to unity, exactly as in the perfectly synchronized case. In order to distinguish between the resting and the synchronized state, one therefore needs to introduce an order parameter that decreases, if the elements collectively slow down. We consider here the well-known order parameter introduced by Kuramoto and Shinomoto \cite{ShiKur86}:
\begin{equation}
\zeta_{\omega,K}(t)=\left|\rho_1(t|\omega,K)-\overline{\rho_1(t|\omega,K)}\right|,
\label{zeta1}
\end{equation}
where $\rho_1(t|\omega,K)=r_{\omega,K}(t)\exp\left[i\Theta_{\omega,K}(t)\right]$ is the first coefficient of the Fourier series expansion of the one-oscillator probability density \eqref{expansion}. 

From now on, if we do not indicate an explicit time-de\-pen\-den\-ce, we refer to long-time averages. In Fig.
\ref{test0}, we show the long-time averaged order parameters for certain sets of parameters, along with the bifurcation values as they can be extracted from Fig. \ref{bif}. The three main
regions mentioned for the bifurcation diagram \ref{bif} can be discriminated here as follows. While the Kuramoto order parameters are close to unity, and the Kuramoto-Shinomoto order parameters are nearly vanishing, the whole system is at rest, and single units do not fire. A partially synchronized oscillation on the global scale is achieved if both the Kuramoto-Shinomoto and the Kuramoto order parameters attain non-zero values. The third region is characterized
by vanishing Kuramoto-Shinomoto and small Kuramoto order parameters. In this case, single units do fire, but in an incoherent way. 
In Fig. \ref{test0}, panels (a) and (c), the humps in the Kuramoto-Shinomoto order parameters signal excitable behavior: for small noise intensities the population stays at rest, then at the SNIC bifurcation (first vertical dash-dotted line) one observes a transition to partial synchronization. Upon further increasing of the noise intensity the population becomes completely incoherent, which happens precisely at the 
Hopf bifurcation (second vertical dash-dotted line). Panels (b) and (d) show no excitable behavior, but only a single transition at the Hopf bifurcation from partial synchronization to incoherence.
Noteworthy however, the Kuramoto-Shinomoto order parameters depend non-monotonically on the noise intensity $D$, such that the highest level of synchronization is achieved at some non-zero noise intensity. 
The standard Kuramoto model cannot uncover this phenomenon.

\begin{figure}
\centering
\includegraphics[width=0.95\linewidth]{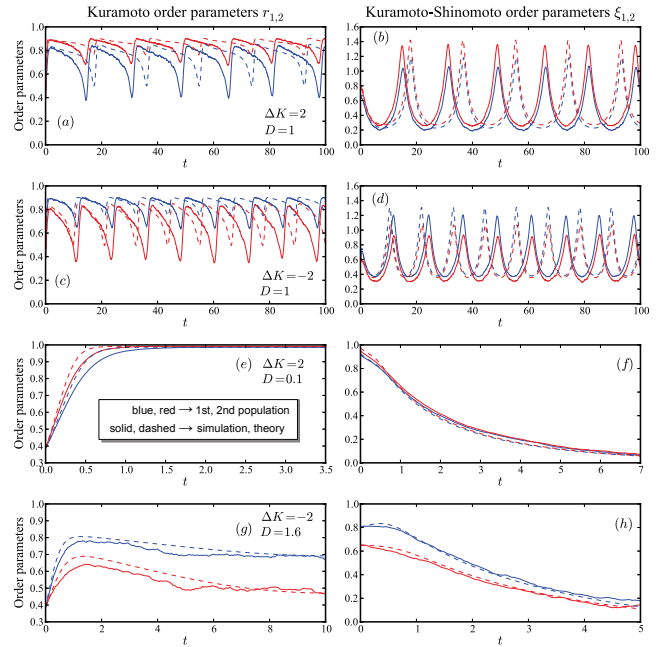}
\caption[]{(Color online) Time-dependent order parameters, theory vs. simulation. The remaining parameters are frequency mismatch $\Delta\omega=0.5$ and average coupling strength $K_0=4$.}
\label{timed}
\end{figure}

Finally, one can observe that the theory agrees very well with the 
results from numerical simulations. Note that the log scale is not necessary to appreciate the accuracy, the latter is chosen in order to emphasize the humps in the excitable regime 
(compare with Ref. \cite{TesScToCo07}). In the appendix \ref{accuracy} we discuss the accuracy in more detail.
The numerical simulations are conducted by integrating the stochastic equations of motion \eqref{ourmodel} using the Heun scheme with time step $0.05$ and considering populations of $N=10^4$ oscillators. Exactly one half of the population is assigned with frequencies and couplings $\left(\omega,K\right)=\left(\omega_1,K_1\right)$, and the second half with $\left(\omega,K\right)=\left(\omega_2,K_2\right)$. Initial conditions of the phases 
$\phi_i(t=0)$ are Gaussian distributed with mean $m(t=0)=0$ and standard deviation $\sigma(t=0) = \sqrt{2}$. Long-time averaged behavior of the order parameters is calculated by averaging the data between 
$t=[2500,5000]$. For the theoretical lines we integrated the reduced system \eqref{mvar} with the same integration parameters.

Figure \ref{timed} shows that even the time-dependent behavior is correctly described by the reduced system \eqref{mvar}. The parameters can be compared with the bifurcation diagram, Fig. \ref{bif}. Note that the Kuramoto-Shinomoto order parameters can exceed unity as a function of time (cf. Sec. \ref{geometric}).
Panels (a)--(d) reflect partially synchronized states, panels (e)--(f) represent resting behavior, and panels (g)--(h) correspond to incoherent dynamics. Apart from time shifts, the qualitative behavior is well predicted by the theory. In the collectively oscillating regime, the theoretical lines lag behind the simulation results for positive coupling mismatch $\Delta K$, but the order is reversed for negative $\Delta K$. Note that the initial values are not perfectly the same as a matter of fact. Finally, figure \ref{timed} illustrates a fundamental feature of the active rotator model, namely the inhomogeneous evolution of the
phases. Such a property results in periodically oscillating mean-field amplitudes and order parameters under partial synchronization, see panels (a)-(d). Moreover, in the incoherent regime the classical Kuramoto order parameter does not vanish, see panel (g).

\section{Conclusion}
\label{conclusion}
In this paper we have studied the active rotator model \cite{ShiKur86} with distributed natural frequencies and coupling 
strengths. The crucial parameter in such excitable systems is the noise intensity \cite{AniAstNeVaSchi07,LiGarNeiSchi04}.
In the infinite system-size limit, we have first derived the exact mean-field description. 
Assuming then that the phases in each set of oscillators with the same natural frequency and coupling strength obey a Gaussian 
distribution with time-dependent cumulants, we have found a representation of the system that permits exemplary scenarios 
composed of a few differential equations. We have used this approach to analyze a mixed population, where one half has been chosen 
to be excitable, whereas the other half  has been in a self-oscillatory state. The distinction depends on whether the natural 
frequency lies below or above the  excitation threshold, respectively. Moreover, the elements of the two subpopulations have 
differed in their individual coupling strengths. In this way we have investigated how frequency and coupling mismatches affect the collective dynamics.
In particular, we have performed a numerical bifurcation analysis in the plane spanned by the noise intensity and the 
frequency mismatch and have shown how these diagrams change as a function of the coupling mismatch. We have found that both 
large frequency and coupling mismatches impede the emergence of synchronized oscillations. This is consistent with
the common finding that oscillatory units which are more distinct, are harder to synchronize. Most intriguingly 
however, we have found that excitability in the whole system is only present, if the excitable elements have a weaker
coupling than the self-oscillatory ones. In other words, a positive coupling-frequency correlation is necessary to cause the excitable behavior in 
the mixed population. We have further found that bistability between various 
collective behaviors is only possible if the positive coupling-frequency correlation is strong enough. Such a 
phenomenon was previously reported only for systems without excitable dynamics, see e.g. Refs. 
\cite{GogaGoArMo11,ZhHuKuLiu13,ChHHuShH13}. The embedded self-oscillatory units considered here can be regarded as 
pacemaker cells in neuronal \cite{Buz06} or cardiac \cite{DiFra93,Ka13,QuHuGarWe14} tissues. Hence, we believe that the work 
presented here contributes to a better understanding of the collective dynamics observed in those systems. Finally, our work may provide a 
new perspective on the emergence of excitable behavior on the global scale, as it is observed e.g. in nonlinear optical cavities \cite{GoMaCo05}. 
It would be interesting to further analyze effects of asymmetries in the natural frequencies and the coupling strengths, as it was 
done e.g. in \cite{AlMi11} for a deterministic system. Moreover, one should also examine the situation where individual 
coupling strengths appear not outside but inside the coupling term, or where the interactions are allowed to be repulsive (see \cite{HoStr12}).

\appendix
\section{The stochastic Kuramoto model with disordered coupling strengths}
\label{KuramotoCouplings}
Here we derive analytical results for the stochastic Kuramoto model (see Refs. \cite{PikRu99,BerGiaPak10,SonnSchi13} and 
\cite{Cha14} for the equivalent Brownian mean-field model) with distributed coupling strengths constituting
the only source of quenched disorder, i.e.
\begin{equation}
  \dot{\phi}_i=\xi_i(t)+\frac{K_i}{N}\sum_{j=1}^{N}\sin\left(\phi_j-\phi_i\right),
\label{kuramoto_d_o_coupling}
\end{equation}
compare with Eq. \ref{ourmodel}. For an interesting recent study of the quenched limit $\xi_i(t)\rightarrow\omega_i$, we refer to \cite{IaPeMcSte13}.

Making use of Eq. \eqref{dotrtheta}, Sec. \ref{gaussian_approx}, we get
\begin{equation}
\begin{cases}
\dot{r}_{K}=&-r_{K}D+\frac{1-r_{K}^4}{2}K\big\langle r_{K'}\cos\left(\Theta_{K'}-\Theta_{K}\right)\big\rangle,\\
\dot{\Theta}_{K}=&\frac{r_{K}^{-1}+r_{K}^3}{2}K\big\langle r_{K'}\sin\left(\Theta_{K'}-\Theta_{K}\right)\big\rangle,
\end{cases}
\label{dotrtheta_d_o_coupling}
\end{equation}
with a single average over the coupling strengths, $\langle\ldots\rangle\equiv\int\mathrm dK'\ldots P(K')$. Applying
the same arguments as in Ref. \cite{SonnSchi13}, one reveals that the critical noise intensity for the transition from
partial synchronization to complete incoherence equals precisely $D_c=\langle K'\rangle/2$.

It is illustrative to come back to the mixed population case with two constituents, i.e.
\begin{equation}
\begin{cases}
\dot{r}_{1}=&-r_{1}D+\frac{1-r_{1}^4}{4}K_1\left[r_1+r_2\cos\left(\Theta_{2}-\Theta_{1}\right)\right],\\
\dot{\Theta}_{1}=&\frac{r_{1}^{-1}+r_{1}^3}{4}K_1 r_{2}\sin\left(\Theta_{2}-\Theta_{1}\right);
\end{cases}
\label{rtheta}
\end{equation}
the equations for $\dot{r}_{2}$ and $\dot{\Theta}_{2}$ are obtained by replacing $1\leftrightarrow 2$.
It is possible to make progress in the stationary regime, $t\rightarrow\infty$, where the derivatives with respect to time
\begin{figure}
\centering
\includegraphics[width=0.7\linewidth]{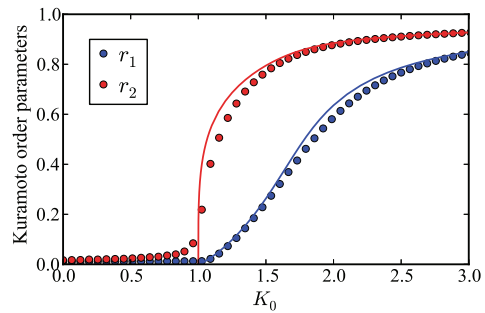}
\caption[]{(Color online) Stationary Kuramoto order parameters for the two subpopulations, theory [lines, cf. Eq. \eqref{r2}] vs. simulation of full dynamics [dots, cf. Eq. \eqref{kuramoto_d_o_coupling}]. Coupling mismatch equals $\Delta K=2$ and noise intensity $D=0.5$.}
\label{DK_2}
\end{figure}
vanish. In order that $\dot{\Theta}_{1}=0$, either $K_1$ or $r_2$ has to vanish or it must hold $\Theta_2=\Theta_1+m\pi,\ m\in\mathbb Z$.
The first two choices need not to be considered, if we are interested in the partially synchronized state. Then from imposing $\dot{r}_{1}=0$
it directly follows that (note that the mean-field amplitude is a non-negative quantity)
\begin{equation}
r_2=r_1\left(\frac{4D}{\left(1-r_1^4\right)K_1}-1\right).
\label{r2}
\end{equation}
The analogous result follows for $r_1$ with the replacement $1\leftrightarrow 2$. This is an interesting result \textit{per se}, 
as it analytically relates the long-time levels of synchronization in the two subpopulations. Note that $r_{1,2}=1$ is achievable only
for $D=0$ or $K_{1,2}\rightarrow\infty$, respectively. The two coupled equations for $r_{1,2}$ \eqref{r2} can be solved 
simultaneously in a numerical way. The results are depicted in Fig. \ref{DK_2}. The order parameter for the first 
subpopulation with the smaller coupling strength shows an anomalous scaling beyond the critical value, clearly different from
the normal square-root scaling; apparently it does not even follow a critical power-law, but rather shows an exponential scaling.
As a consequence, slightly above the critical coupling, one observes chimera-like states, where one subpopulation shows significant 
synchronization, while at the same time the other one stays almost incoherent. This scenario is reminiscent of what has been found in 
Refs. \cite{AbMiStWi08,La09}. We expect this to be a promising direction for future studies.

\section{A geometric view of the Kuramoto-Shinomoto order parameter}
\label{geometric}
\begin{figure}
\centering
\includegraphics[width=0.3\linewidth]{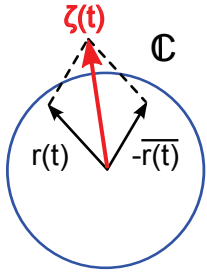}
\caption[]{(Color online) The unit circle in the complex plane helps to visualize the order parameters considered here, see Eq. \ref{zeta}.
Vectors are denoted by their absolute values.}
\label{shino_fig}
\end{figure}

Let us first repeat the definition of the Kuramoto-Shinomoto order parameter (for simplicity, we neglect here the subdivision into distinct sets of natural frequencies and couplings):
\begin{equation}
\zeta(t)=\left|\rho_1(t)-\overline{\rho_1(t)}\right|,
\label{zeta}
\end{equation}
where $\rho_1(t)=r(t)\exp\left[i\Theta(t)\right]$. Now in the complex plane, $\zeta(t)$ corresponds to the length
of one of the diagonals in the parallelogram that is
spanned by the vectors $\rho_1(t)$ and $-\overline{\rho_1(t)}$. This is visualized in Fig. \ref{shino_fig}, where for 
simplicity the vectors are denoted by their absolute values. Clearly, $\zeta(t)$ can be larger than unity. In fact, 
the maximal value is $2$. However, this can be true only in an infinitely small time period, because $\overline{\rho_1(t)}$ 
is the long-time average of $\rho_1(t)$. To make this point more illustrative, imagine that half of the time the Kuramoto
order parameter $\rho_1(t)$ is given by some unit vector, and the other half of the time by the zero vector. Then the 
Kuramoto-Shinomoto order parameter $\zeta(t)$ will equal $1/2$ at all times, see Eq. \ref{zeta} and Fig. \ref{shino_fig}. 
It is straightforward to see in this manner that the long-time averaged Kuramoto-Shinomoto order parameter always lies between 
zero and unity.

\section{On the accuracy of the Gaussian approximation (GA)}
\label{accuracy}
\begin{figure}
\centering
\includegraphics[width=0.95\linewidth]{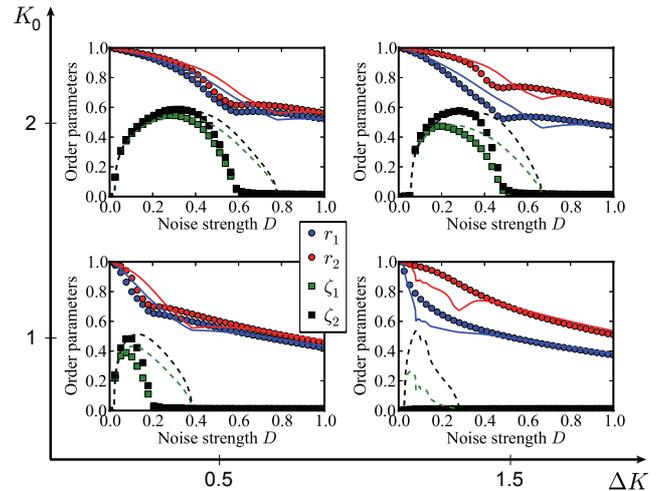}
\caption[]{(Color online) Long-time averaged order parameters as a function of the noise intensity are compared for different average coupling strengths $K_0$ and
coupling mismatches $\Delta K$. All lines correspond to theoretical results, cf. Eq. \eqref{mvar}, while the symbols result from a simulation of the full dynamics \eqref{ourmodel}. The frequency mismatch is fixed at $\Delta\omega=0.1$.}
\label{test1}
\end{figure}

It is well-known that a small noise intensity favors the GA, see e.g. \cite{KurSchu95}.
In the recent paper \cite{SonnSchi13} a systematic examination of the GA was carried out for 
the stochastic Kuramoto model. It was found that the critical coupling strength for the onset of synchronization
is exactly recovered. Also below and sufficiently above (twice as much) the critical value, the GA is highly accurate. 
Here we proceed to show that a large coupling mismatch deteriorates the accuracy
of the GA, in particular if the average coupling strength is small. In fact, the combination of large coupling
mismatch $\Delta K$ and small average coupling strength $K_0$ is the only case where we find qualitative disagreement.
Large noise intensities merely decrease the quantitative agreement and the frequency mismatch alone does not cause any 
inaccuracies. In Fig. \ref{test1} we depict the appearance of the qualitative disagreement for large $\Delta K$, but small $K_0$.
There, the theory predicts excitable behavior, which is not reproduced by numerical simulations of the full system.
The disagreement seems to be accompanied by an additional wiggle in the theoretical curves. Furthermore, one can clearly
see the increased quantitative discrepancy for larger noise intensities $D$. Correspondingly, the saddle-node bifurcation
is in general better reflected by the theory than the Hopf bifurcation in the excitable system. For very large noise
intensities, i.e. in the incoherent state, the agreement between theory and simulations is recovered.

\begin{acknowledgement}
B.S. thanks M. A. Zaks for helpful discussions. B.S. further acknowledges support
from the Deutsche Forschungsgemeinschaft (GRK1589/1). L.SG. acknowledges the Bernstein Center for 
Computational Neuroscience Berlin (project A3). T.K.DM.P. acknowledges FAPESP (grant 2012/22160-7) and IRTG 1740. F.A.R. acknowledges CNPq (grant 305940/2010-4), Fapesp (grant \-2013/26416-9) and IRTG 1740 (DFG and FAPESP) for financial support. J.K. acknowledges IRTG 1740 (DFG and FAPESP) for the sponsorship provided.
\end{acknowledgement}

\bibliography{bibliography}

\begin{thebibliography}{10}

\bibitem{Gl01}
L.~Glass.
\newblock {\em Nature}, 410:277--284, 2001.

\bibitem{Buz06}
G.~Buzs\'aki.
\newblock {\em {Rhythms of the Brain}}.
\newblock Oxford University Press, 2006.

\bibitem{PoPoLSG12}
D.~E. Postnov, D.~D. Postnov, and L.~Schimansky-Geier.
\newblock {\em Brain research}, 1434:200--211, 2012.

\bibitem{LuBaSo13}
T.~B. Luke, E.~Barreto, and P.~So.
\newblock {\em Neural Comput.}, 25(12):3207--3234, 2013.

\bibitem{LuBenBaBoTo14}
S.~Luccioli, E.~Ben-Jacob, A.~Barzilai, P.~Bonifazi, and A.~Torcini.
\newblock {\em PLoS Comput. Biol. (submitted)}, 2014.

\bibitem{DiFra93}
D.~DiFrancesco.
\newblock {\em Annu. Rev. Physiol.}, 55(1):455--472, 1993.

\bibitem{ShaBoShGl13}
T.~K. Shajahan, B.~Borek, A.~Shrier, and L.~Glass.
\newblock {\em New J. Phys.}, 15(2):023028, 2013.

\bibitem{Ka13}
A.~Karma.
\newblock {\em Annu. Rev. Condens. Matter Phys.}, 4(1):313--337, 2013.

\bibitem{QuHuGarWe14}
Z.~Qu, G.~Hu, A.~Garfinkel, and J.~N. Weiss.
\newblock {\em Phys. Rep. (in press)}, 2014.

\bibitem{Br08}
M.~Brede.
\newblock {\em Phys. Lett. A}, 372:2618, 2008.

\bibitem{GogaGoArMo11}
J.~G\'omez-Gardenes, S.~G\'omez, A.~Arenas, and Y.~Moreno.
\newblock {\em Phys. Rev. Lett.}, 106:128701, 2011.

\bibitem{peron2012determination}
T.~K.~DM. Peron and F.~A. Rodrigues.
\newblock {\em Phys. Rev. E}, 86:056108, 2012.

\bibitem{leyva2012explosive}
I.~Leyva, R.~Sevilla-Escoboza, J.~M. Buld\'u, I.~Sendi{\~n}a-Nadal,
  J.~G\'omez-Garde{\~n}es, A.~Arenas, Y.~Moreno, S.~G\'omez,
  R.~Jaimes-Re\'ategui, and S.~Boccaletti.
\newblock {\em Phys. Rev. Lett.}, 108:168702, 2012.

\bibitem{coutinho2013kuramoto}
B.~C. Coutinho, A.~V. Goltsev, S.~N. Dorogovtsev, and J.~F.~F. Mendes.
\newblock {\em Phys. Rev. E}, 87:032106, 2013.

\bibitem{SonnSagSchi13}
B.~Sonnenschein, F.~Sagu\'es, and L.~Schimansky-Geier.
\newblock {\em Eur. Phys. J. B}, 86:12, 2013.

\bibitem{SkSuTaRes13}
P.~S. Skardal, J.~Sun, D.~Taylor, and J.~G. Restrepo.
\newblock {\em Eur. Phys. Lett.}, 101:20001, 2013.

\bibitem{li2013reexamination}
P.~Li, K.~Zhang, X.~Xu, J.~Zhang, and M.~Small.
\newblock {\em Phys. Rev. E}, 87:042803, 2013.

\bibitem{SuRuGuLi13}
G.~Su, Z.~Ruan, S.~Guan, and Z.~Liu.
\newblock {\em Europhys. Lett.}, 103(4):48004, 2013.

\bibitem{JiPeMeRodKu13}
P.~Ji, T.~K.~DM. Peron, P.~J. Menck, F.~A. Rodrigues, and J.~Kurths.
\newblock {\em Phys. Rev. Lett.}, 110:218701, 2013.

\bibitem{zhu2013criterion}
L.~Zhu, L.~Tian, and D.~Shi.
\newblock {\em Phys. Rev. E}, 88:042921, 2013.

\bibitem{ZhHuKuLiu13}
X.~Zhang, X.~Hu, J.~Kurths, and Z.~Liu.
\newblock {\em Phys. Rev. E}, 88:010802(R), 2013.

\bibitem{LeNaSenAlmBuZaPaBoc13}
I.~Leyva, A.~Navas, I.~Sendi{\~n}a-Nadal, J.~A. Almendral, J.~M. Buld\'u,
  M.~Zanin, D.~Papo, and S.~Boccaletti.
\newblock {\em Sci. Rep.}, 3:1281, 2013.

\bibitem{zhu2013explosive}
L.~Zhu, L.~Tian, and D.~Shi.
\newblock {\em Eur. Phys. J. B}, 86:451, 2013.

\bibitem{ChHHuShH13}
H.~Chen, G.~He, F.~Huang, C.~Shen, and Z.~Hou.
\newblock {\em Chaos}, 23(3):033124, 2013.

\bibitem{ZoPerSmLiKu14}
Y.~Zou, T.~Pereira, M.~Small, Z.~Liu, and J.~Kurths.
\newblock {\em Phys. Rev. Lett.}, 112:114102, 2014.

\bibitem{SkAr14}
P.~S. Skardal and A.~Arenas.
\newblock {\em Phys. Rev. E}, 89:062811, 2014.

\bibitem{mikkelsen2013emergence}
K.~Mikkelsen, A.~Imparato, and A.~Torcini.
\newblock {\em Phys. Rev. Lett.}, 110:208101, 2013.

\bibitem{orlandi2013noise}
J.~G. Orlandi, J.~Soriano, E.~Alvarez-Lacalle, S.~Teller, and J.~Casademunt.
\newblock {\em Nat. Phys.}, 9(9):582--590, 2013.

\bibitem{ShiKur86}
S.~Shinomoto and Y.~Kuramoto.
\newblock {\em Prog. Theor. Phys.}, 75(5):1105, 1986.

\bibitem{KurSchu95}
C.~Kurrer and K.~Schulten.
\newblock {\em Phys. Rev. E}, 51(6):6213, 1995.

\bibitem{ZaNeFeSch03}
M.~A. Zaks, A.~B. Neiman, S.~Feistel, and L.~Schimansky-Geier.
\newblock {\em Phys. Rev. E}, 68:066206, 2003.

\bibitem{SonnZaNeiLSG13}
B.~Sonnenschein, M.~A. Zaks, A.~B. Neiman, and L.~Schimansky-Geier.
\newblock {\em Eur. Phys. J. Special Topics}, 222:2517, 2013.

\bibitem{SonnSchi13}
B.~Sonnenschein and L.~Schimansky-Geier.
\newblock {\em Phys. Rev. E}, 88:052111, 2013.

\bibitem{TaPak01}
S.~Tanabe and K.~Pakdaman.
\newblock {\em Phys. Rev. E}, 63:031911, 2001.

\bibitem{ZaSaLSGNei05}
M.~A. Zaks, X.~Sailer, L.~Schimansky-Geier, and A.~B. Neiman.
\newblock {\em Chaos}, 15:026117, 2005.

\bibitem{Bu01}
A.~N. Burkitt.
\newblock {\em Biol. Cybern.}, 85:247--255, 2001.

\bibitem{LafTo10}
L.~F. Lafuerza and R.~Toral.
\newblock {\em J. Stat. Phys.}, 140:917--933, 2010.

\bibitem{BuRanTodVas10}
N.~Buri\'c, D.~Rankovi\'c, K.~Todorovi\'c, and N.~Vasovi\'c.
\newblock {\em Physica A}, 389:3956--3964, 2010.

\bibitem{FraTodVasBu13}
I.~Franovi\'c, K.~Todorovi\'c, N.~Vasovi\'c, and N.~Buri\'c.
\newblock {\em Phys. Rev. E}, 87:012922, 2013.

\bibitem{FraTodVasBu14}
I.~Franovi\'c, K.~Todorovi\'c, N.~Vasovi\'c, and N.~Buri\'c.
\newblock {\em Phys. Rev. E}, 89:022926, 2014.

\bibitem{AniAstNeVaSchi07}
V.~S. Anishchenko, V.~Astakhov, A.~Neiman, T.~Vadivasova, and
  L.~Schimansky-Geier.
\newblock {\em Nonlinear Dynamics of Chaotic and Stochastic Systems}.
\newblock Springer-Verlag, Berlin, 2007.

\bibitem{LiGarNeiSchi04}
B.~Lindner, J.~Garc\'ia-Ojalvo, A.~Neiman, and L.~Schimansky-Geier.
\newblock {\em Phys. Rep.}, 392:321, 2004.

\bibitem{Ris96}
H.~Risken.
\newblock {\em {The Fokker-Planck Equation}}.
\newblock Springer, 1996.

\bibitem{faugeras2009constructive}
O.~Faugeras, J.~Touboul, and B.~Cessac.
\newblock {\em Front. Comput. Neurosci.}, 3:1, 2009.

\bibitem{mischler2013kac}
S.~Mischler and C.~Mouhot.
\newblock {\em Invent. math.}, 193(1):1--147, 2013.

\bibitem{kac1954foundations}
M.~Kac.
\newblock Foundations of kinetic theory.
\newblock In {\em Proc. Third Berkeley Symp. on Math. Statist. and Prob.},
  volume~3, pages 171--197. Univ. of Calif. Press, 1956.

\bibitem{bonilla1993glassy}
L.~L. Bonilla, C.~J.~P{\'e}rez Vicente, and J.~M. Rubi.
\newblock {\em J. Stat. Phys.}, 70(3-4):921--937, 1993.

\bibitem{HilBuCh07}
E.~J. Hildebrand, M.~A. Buice, and C.~C. Chow.
\newblock {\em Phys. Rev. Lett.}, 98:054101, 2007.

\bibitem{SonnSchi12}
B.~Sonnenschein and L.~Schimansky-Geier.
\newblock {\em Phys. Rev. E}, 85:051116, 2012.

\bibitem{Fr05}
T.~D. Frank.
\newblock {\em Nonlinear Fokker-Planck equations: fundamentals and
  applications}.
\newblock Springer, 2005.

\bibitem{OttAnt08}
E.~Ott and T.~M. Antonsen.
\newblock {\em Chaos}, 18:037113, 2008.

\bibitem{MoKuBl04}
E.~Montbri\'o, J.~Kurths, and B.~Blasius.
\newblock {\em Phys. Rev. E}, 70:056125, 2004.

\bibitem{AbMiStWi08}
D.~M. Abrams, R.~Mirollo, S.~H. Strogatz, and D.~A. Wiley.
\newblock {\em Phys. Rev. Lett.}, 101:084103, 2008.

\bibitem{La09}
C.~R. Laing.
\newblock {\em Chaos}, 19:013113, 2009.

\bibitem{HoStr11PRL}
H.~Hong and S.~H. Strogatz.
\newblock {\em Phys. Rev. Lett.}, 106:054102, 2011.

\bibitem{HoStr11}
H.~Hong and S.~H. Strogatz.
\newblock {\em Phys. Rev. E}, 84:046202, 2011.

\bibitem{PaMo14}
D.~Paz\'o and E.~Montbri\'o.
\newblock {\em Phys. Rev. X}, 4:011009, 2014.

\bibitem{DaNak04}
H.~Daido and K.~Nakanishi.
\newblock {\em Phys. Rev. Lett.}, 93(10):104101, 2004.

\bibitem{PaMon06}
D.~Paz\'o and E.~Montbri\'o.
\newblock {\em Phys. Rev. E}, 73:055202(R), 2006.

\bibitem{DaKaNi13}
H.~Daido, A.~Kasama, and K.~Nishio.
\newblock {\em Phys. Rev. E}, 88:052907, 2013.

\bibitem{AlMi11}
L.~M. Alonso and G.~B. Mindlin.
\newblock {\em Chaos}, 21:023102, 2011.

\bibitem{DhGoKuz03}
A.~Dhooge, W.~Govaerts, and Y.~A. Kuznetsov.
\newblock {\em ACM TOMS}, 29:141--164, 2003.

\bibitem{TesScToCo07}
C.~J. Tessone, A.~Scir\`e, R.~Toral, and P.~Colet.
\newblock {\em Phys. Rev. E}, 75:016203, 2007.

\bibitem{GoMaCo05}
D.~Gomila, M.~A. Mat\'ias, and P.~Colet.
\newblock {\em Phys. Rev. Lett.}, 94:063905, 2005.

\bibitem{HoStr12}
H.~Hong and S.~H. Strogatz.
\newblock {\em Phys. Rev. E}, 85:056210, 2012.

\bibitem{PikRu99}
A.~Pikovsky and S.~Ruffo.
\newblock {\em Phys. Rev. E}, 59(2):1633, 1999.

\bibitem{BerGiaPak10}
L.~Bertini, G.~Giacomin, and K.~Pakdaman.
\newblock {\em J. Stat. Phys.}, 138:270--290, 2010.

\bibitem{Cha14}
P.-H. Chavanis.
\newblock {\em Eur. Phys. J. B}, 87:120, 2014.

\bibitem{IaPeMcSte13}
D.~Iatsenko, S.~Petkoski, P.~V.~E. McClintock, and A.~Stefanovska.
\newblock {\em Phys. Rev. Lett.}, 110:064101, 2013.

\end{thebibliography}
\bibliographystyle{unsrt}

\end{document}